\documentstyle[aps,preprint]{revtex}
\begin{document}

\draft

\title{Stochastic variational method with noncentral forces}

\author{K. Varga$^{1,2}$, Y. Ohbayasi$^3$ and Y. Suzuki$^2$}
\address{$^1$Institute of Nuclear Research, Hungarian Academy of Sciences,
(ATOMKI),
H-4001 Debrecen, Hungary\\
$^2$Department of Physics, Niigata University
Niigata 950-21, Japan\\
$^3$Graduate School of Science and Technology, Niigata University
Niigata 950-21, Japan}

\date{\today}

\maketitle

\begin{abstract}
This paper extends the application of the stochastic variational
method to noncentral interactions. Several examples are presented 
for three- and four-nucleon systems with realistic nuclear forces.
The correlated Gaussians easily cope with the strong short
range repulsion of the potential and the stochastic variational
method efficiently selects the most important spin, isospin and
orbital angular momentum components of the wave function. The examples confirm 
the usefulness and accuracy of the method.
\end{abstract}

\narrowtext

In a recent paper \cite{VS} we have shown that the stochastic variational
method with correlated Gaussian basis (SVM) provides an accurate solution 
for various few-body systems.
The applications of the SVM, however, were limited to central interactions
so far \cite{VS,VSL,Kuk}. The real challenge in nuclear few-body problems is the solution 
with a realistic nucleon-nucleon interaction including spin-orbit, 
tensor, etc. potentials. The application of the SVM for realistic 
interactions is,  
therefore, a stringent test of the usefulness of the method.
\par\indent
The SVM selects the  most appropriate basis
states in a trial and error procedure: various randomly generated trial 
states are tested and the usefulness of these states are judged from 
their contribution to the energy of the system. As it is a variational 
method, the  choice of the functional form of the trial states greatly
determines the accuracy and applicability of the method. We prefer the
 correlated Gaussians as trial functions, as these functions proved to 
be enormously accurate in few-body problems with zero orbital angular momentum. 
It is remarkable that the variational calculations with  these functions 
give one of the most precise results up to date, for example, 
for the hydrogen molecular ion, the helium atom and the positronium molecule,  
reaching 13 digits accuracy \cite{cen}. Another advantage of the 
correlated Gaussian functions is that one can easily calculate their 
matrix elements \cite{VS}
and one can easily extend the calculation to more than A=3,4 particle
systems. These features make it especially interesting to test their
applicability in few-nucleon systems with realistic forces.
\par\indent
The secret  of  the accuracy of the variational methods on the correlated  
Gaussian basis is the careful selection (``optimization'') 
of the nonlinear parameters. In the SVM with central interaction these 
nonlinear parameters are the subject of the random test. 
\par\indent
In nuclear physics the spin, isospin and orbital angular momentum 
dependence of the 
interaction is extremely important. To extend the application of the SVM
for nuclear systems with realistic interactions,  
besides the previously applied  random selection
of the nonlinear parameters of the correlated Gaussians,
a random selection of the  spin and orbital components of the wave 
function is introduced.
\par\indent
The literature is very rich in papers devoted to  few-nucleon problems 
\cite{GFMC1,GFMC2,Car,CHH,Kam,FY,FY2,Friar,A6,VMC1,Aka}. For A=3 nuclei it is possible to 
include enough channels and an exact Faddeev calculation can be performed 
\cite{Friar}, but other methods
(e.g., Variational Monte Carlo \cite{VMC1}, Green Function Monte Carlo 
\cite{GFMC1}, Correlated Hyperspherical Harmonics \cite{CHH}, 
ATMS \cite{Aka}, or the Gaussian Variational Method (GVM) \cite{Kam}) give 
essentially the same results. For A=4 nuclei some of these methods become 
too complicated, and the results of the existing calculations 
\cite{GFMC1,CHH,FY,VMC1} agree only within a few hundreds keV. The development 
of other methods is therefore very important. 
\par\indent
The purpose of this paper is to 
present a SVM calculation for the $^3$H and $^4$He nuclei 
with the interactions V6
(1, $\sigma\cdot\sigma$, $\tau\cdot\tau$, $\sigma\cdot\sigma \tau\cdot\tau$,
 S, S$\tau\cdot\tau$) and  V8 (1, $\sigma\cdot\sigma$, $\tau\cdot\tau$, 
$\sigma\cdot\sigma \tau\cdot\tau$, S, S$\tau\cdot\tau$, $L\cdot S$, 
$L\cdot S$$\tau\cdot\tau$) and to compare with
the results of other calculations in order to confirm the usefulness 
and accuracy of the SVM.  
\par\indent
The trial function is defined in the following way:
\begin{equation}
\psi_{(LS)JMTM_T}({\bf x},A)={\cal A}\lbrace {\rm e}^{-{1\over 2}
{\bf x} A {\bf x}}\left[\theta_L({\bf x})\chi_S\right]_{JM} \eta_{TM_T}
\rbrace ,
\end{equation}
where ${\bf x}=({\bf x}_1,...,{\bf x}_{N-1})$ is a set of relative (Jacobi)
coordinates, the operator ${\cal A}$ is an antisymmetrizer, 
$\chi_{SM_S}$  is the spin function, and $\eta_{TM_T}$ is the isospin 
function.
The matrix $A$ is an $(A-1)\times(A-1)$ matrix of the nonlinear variational 
parameters.
The function
$\theta_{LM_L}(\bf x)$ represents the angular part of the wave function. It 
is taken as a vector coupled product of partial waves
\begin{equation}
\theta_{LM_L}({\bf x})=\left[\left[\left[
Y_{l_1}({\bf x_1}) Y_{l_2}({\bf x_2})\right]_{l_{12}}
Y_{l_3}({\bf x_3})\right]...\right]_{LM_L}.
\end{equation}
The spin and isospin functions are also successively coupled, for example
\begin{equation}
\chi_{SM_S}=\chi_{s_{12}s_{123}...SM_S}=
[[[\chi_{1\over 2}\chi_{1\over 2}]_{s_{12}}
\chi_{1\over 2}]_{s_{123}}...]_{SM_S}
\end{equation}
For the triton, we used the partial wave channels with $(l_1,l_2)L,\ \ 
(l_1+l_2\le 4,\ \ l_i\le 2)$ and $L=0,1,2$.  For the alpha particle, 
all partial waves
$((l_1,l_2)l_{12},l_3)L$ satisfying the condition $l_1+l_2+l_3\le 4,\ \  
l_i\le 2, \ \ \vert l_1-l_2\vert\le l_{12}\le l_1+l_2$ and $L=0,1,2$ have
been tried. 
There are 3 spin and 2 isospin channels for the triton, and 
6 spin and 2 isospin channels for the alpha particle. 
These channels are listed in Table II of ref. [1].
\par\indent
The basis setup of the SVM can be briefly described 
as follows :
Let us assume that the dimension of the basis is $K-1$.
\par\noindent
(1) Generate ${\cal N}$ random candidates
to find the $K$th basis function:
\par\noindent
(1a) Pick up a spin, isospin and partial wave channel randomly.
(Some of the possibilities are listed in Table I.)
\par\noindent
(1b) Select the nonlinear parameters randomly from a ``physically''
important interval.
\par\noindent
(2) Calculate the ground state energy on the $K$ basis states.
\par\noindent
(3) Select the basis state which gave the lowest energy amongst the
randomly generated trial functions and add it to the basis.
\par\noindent
(4) Increase the dimension to $K+1$.
\par\indent
In principle it would be more powerful to change all the  $K$ basis states
randomly, but that  requires  repetitions of diagonalization of $K\times
K$ matrices and becomes prohibitively computer time consuming. In the procedure
(1)-(2)-(3)-(4) only one of the row (column) of the matrices changes, and 
there is no need for diagonalization to calculate the lowest eigenvalue [1].
A numerical optimatization of the basis would be an other option, but it 
is highly nontrivial. The reason is partly  that  number of 
nonlinear parameters is too large and partly that the channels 
(components of the wave functions with different quantum numbers) cannot be 
treated as continouos parameters and therefore the strategies of the 
numerical optimatizations can hardly be applied. 
\par\indent
The main motivation for the random selection is that the number of channels
and the nonlinear parameters are prohibitively large, therefore 
the calculation of all of the matrix elements and diagonalization 
including all potentially important basis states 
is out of the question. In the case of the 
alpha particle, for example, there are 12 spin-isospin channels and  32 
partial wave channels even in our truncated partial wave
expansion. The simplest choice of the nonlinear
parameters of the Gaussian basis is to use only a diagonal matrix $A$. 
One may choose the diagonal elements as a geometric progression 
\cite{Kam}, for example. To reach good accuracy at least five terms
for each of $A_{11},A_{22}$ and $A_{33}$ have to be used, 
requiring $5^3=125$ functions in a given channel. The spin-isospin 
and space part, therefore, would result in a basis size of about ten 
thousands ($2\times125\times12\times32$) in such a ``direct'' 
calculation for the alpha particle. 
\par\indent
It is difficult to decide which channels are important and how many spatial
functions are needed amongst the above mentioned numerous possibilities. 
The SVM automatically selects the important basis states through steps 
(1)-(2)-(3)-(4). The energy of the already selected states can be improved
further: After reaching a given dimension, say $K$, one can stop the 
calculation and refine the nonlinear parameters.
This fine tuning is done by repeating the steps (1b)-(2)-(3), that is, by 
changing the nonlinear parameters of one of the basis states which belongs
to the already selected basis set. The nonlinear parameters are changed 
in the same spirit as before by randomly selecting the best. 
This is repeated for different basis states until the energy gain is less
than 0.005 MeV. In this case we do not change the channels that are
already selected and keep the dimension fixed. In the first 
stage of the calculation (steps (1)-(2)-(3)-(4)), the first $K-1$ basis 
states are kept frozen, and the best $K$th basis state is selected 
with respect to them, so its selection is limited by  
the already chosen basis states. In the refining steps all of 
these states play active role again. 
Depending on the value of ${\cal N}$, these steps  considerably 
improve the energy. 
The dimensions at the refining steps and the best energies obtained 
on these basis sizes are listed in Tables II and III.
\par\indent
The matrix elements were calculated as described in ref. [1]. To check 
the calculation we compared the numerical values of our matrix elements
to those calculated by Kamimura 
\cite{Kam}. The matrix elements agree in all digits.
\par\indent
To test the method we used two, from the point of view of
spatial form, rather different interactions. The first, the Argonne
potential (AV6 and AV8) \cite{wir}, is a well-behaved smooth function. The second, 
the Reid potential (RV8) \cite{reid} is more singular and includes  a linear
 combination of 
${\rm exp}\lbrace -\mu r^2\rbrace/r^k$-type terms.
The latter potential can certainly cause some problems. The AV6 potential
includes only central and tensor components and serves as a good test case
for the inclusion of noncentral forces. The AV6 potential has the same 
central and tensor components as the AV14 potential \cite{wir}. The AV6
and AV8 potentials are used without Coulomb potential, while in the case 
of the RV8 potential the Coulomb interaction is added.
\par\indent
To keep the number of nonlinear parameters low, we restricted 
the matrix $A$ to be diagonal in one of the possible Jacobi-coordinate 
system. In the case of the three-nucleon system we have only one 
possibility of the Jacobi coordinate ($(NN)+N$), and for the alpha particle
we will use $(3N)+N$ (``K'') (and $(2N)+(2N)$-type  (``H'') 
Jacobi-coordinates. (Note that 
the basis function is fully antisymmetrized.) 
This choice is also dictated by physical intuitions as it is natural
to consider $^3$H+$p$, $^3$He+$n$ and $d+d$-type partitions in the 
alpha particle. Test calculations show that this choice already gives 
satisfactory results.
\par\indent
As seen in Tables II and III,  the convergence is relatively fast
both for the triton and alpha particle with the AV6 and AV8 potentials, 
but in the case of the RV8 potential, due to its
singular nature and stronger repulsive core, the convergence is
slower. The result of the calculation
does not improve by increasing the basis size further. In the case of
the triton higher partial 
waves are not expected to give substantial contribution. In ref.\cite{Kam} 
the authors included $l_1+l_2 \le 4$ partial
waves for the AV14 potential but the energy changed by merely 
0.002 MeV. 
\par\indent
As the different channels are nonorthogonal, it is not trivial to calculate
their weights. As a rough guide for their relative ``importance'',  we list 
the most often selected channels in Table I. By using only these channels
one gets about 0.5 MeV less binding for the alpha particle, so these channels
alone already give a good approximation for the wave function. 
\par\indent
The results of SVM and other calculations are compared in Tables IV and
V. We show and compare the results not only for the energy but for 
the root mean square radius and average kinetic and potential energy.
The results  remain the same by repeating the calculations several 
times starting from different random values. 
The results of SVM and GFMC are very close to each other. Except for
the case of the alpha particle with AV6, the results of SVM are always 
close to the GFMC, though the expectation
values of the kinetic and the potential energies are somewhat different. 
\par\indent
One can get good solution on relatively small basis dimensions. About 50 
basis states for the triton and about 200 basis states for the alpha 
particle give fairly good binding energies. The energy of the alpha 
particle in the basis size of 100 is already within 0.5 MeV of its 
final value. 
\par\indent
Variational calculations were considered to be inappropriate to give
accurate ground state energies for light nuclei, as they might not be
able to reproduce the large cancellation between the kinetic and potential
energy \cite{Car}. Our calculation shows that, with the careful optimization
of the nonlinear variational parameters, this is not the case. We can obtain
accurate energies even in the case of the RV8 potential. 
We note that the $L\cdot S$ term of the RV8 potential has been
cut off at magnitude of 1 GeV in the GFMC calculation to reduce the
statistical fluctuations caused by the 1/r singularity of the potential. 
To be consistent with this,
we repeated the calculation with and without this cut off. 
We got 0.02 MeV less binding for the alpha particle with the modified 
potential, so the difference is essentially negligible.
\par\indent
Once we have a basis for a given potential, one naturally expects
that the same basis may give fair ground state energy for other potentials
of similar nature as well. We have checked if it is really the case. 
The basis optimized 
for the RV8 potential gives excellent results (within 100 keV) for 
AV6 and AV8 as well. Due to the singular nature and stronger
repulsive core of RV8, however, the basis optimized for AV8 gives
about 500 keV less binding for the alpha particle with RV8 than the basis 
optimized for RV8 itself. This result is still not so bad and can be 
easily improved by refining the nonlinear parameters. Therefore, one does 
not have to look for a new basis set for different interactions, but the 
same basis can be used for a given system with some ``fine tuning'' if 
necessary.
The same is hopefully true for the other ($(LS)^2,L^2$, etc.) part of the 
interaction: an already selected basis can be tailored  to these
additional terms, that is, instead of the steps (1)-(2)-(3)-(4),  
the several times faster refining steps (1b)-(2)-(3) can be used. 
\par\indent
It is obviously important to calculate accurate binding energies of 
light nuclei with realistic forces. For example, the two-neutron separation 
energy of $^6$He is about 1 MeV, so that a few hundreds keV less binding
is thought to change its neutron ``halo'' structure significantly.
One of the advantages of the SVM is that it is relatively easy to extend
it to A=5, 6, 7 nucleon systems [1]. The low dimension of the bases required 
to solve the A=3,4 nucleon problems confirms that it is possible to 
treat larger nuclei with realistic forces with the SVM. 
The formalism and the computer code itself is general [1], so the 
applicability is mostly limited by the memory and speed of the available 
computer. 
The partial wave decomposition would not be appropriate for larger systems, 
but can be efficiently substituted by a simpler formalism [1]. This study is 
under way. 
\par\indent
In summary, we have presented a stochastic variational solution
for the triton and alpha particle with realistic nuclear forces. 
The wave function of these systems is a linear combination of components
with different spin, isospin, orbital angular momentum and  space parts. 
To find a good variational solution, one has to choose each part of the
components with due care. The stochastic variational method attempts to 
collect the most important basis functions by randomly selecting the 
spin, isospin, orbital angular momentum and  space parts.  
The fact that the results are in good agreement with those of the best
calculations in the literature encourages the applications of the method
for other cases, such as A=6 and 7 nuclei. 

\par\indent
The authors are grateful to Prof. M. Kamimura for providing his matrix
elements for comparison, and to Prof. J. Carlson for providing the parameters of 
the potentials and unpublished results. 
This work was supported by OTKA grant No. T17298 (Hungary) and 
by Grants-in-Aid for Scientific Research (No. 05243102 and No. 06640381) 
and for International Scientific Research (Joint Research) (No. 08044065) 
of the Ministry of Education, Science and Culture (Japan). K. V. gratefully
acknowledges the support of JSPS. The authors 
are grateful for the use of RIKEN's VPP500 computer and JAERI's VPP300 
computer which 
made possible most of the calculations.

\begin{table}
\caption{The most frequently selected channels of the three- and 
four-nucleon systems.}
\begin{tabular}{cccc}
$ (l_{1},l_{2})L$  & $s_{12}$ & $S$ & $t_{12}$  \\
\hline
(0,0)0             &     0    & 1/2 & 1       \\
(0,0)0             &     1    & 1/2 & 0       \\
(2,0)2             &     1    & 3/2 & 0       \\
(0,2)2             &     1    & 3/2 & 0       \\
(2,2)0             &     1    & 1/2 & 0       
\end{tabular}
\begin{tabular}{cccccc}
$((l_{1},l_{2})l_{12},l_{3})L$&$s_{12}$&$s_{123}$&$S$&$t_{12}$&Jacobi \\
\hline
((0,0)0,0)0                   & 0      &1/2      &0  & 1      &  H    \\
((0,0)0,0)0                   & 1      &1/2      &0  & 0      &  H    \\
((0,0)0,0)0                   & 0      &1/2      &0  & 1      &  K    \\
((2,2)0,0)0                   & 1      &1/2      &0  & 0      &  K    \\
((2,2)0,0)0                   & 1      &1/2      &0  & 0      &  H    \\
((0,2)2,0)2                   & 1      &3/2      &2  & 0      &  K    \\
((0,2)2,0)2                   & 1      &3/2      &2  & 0      &  H    \\
((2,0)2,0)2                   & 1      &3/2      &2  & 0      &  H    \\
((1,1)2,0)2                   & 1      &3/2      &2  & 1      &  K    \\
((2,2)2,0)2                   & 1      &3/2      &2  & 0      &  K    
\end{tabular}
\end{table}

\begin{table}
\caption{The convergence of energy (in MeV) for the triton.}
\begin{tabular}{cccc}
$ K $   &   AV6     &  AV8     & RV8    \\
\hline
 25      &  $-6.63$    & $-7.36$  & $-6.53$ \\
 50      &  $-7.04$    & $-7.69$  & $-7.41$ \\
 75      &  $-7.11$    & $-7.74$  & $-7.54$ \\
100      &  $-7.15$    & $-7.79$  & $-7.59$       
\end{tabular}
\end{table}
\begin{table}
\caption{The convergence of energy (in MeV) for the alpha particle.}
\begin{tabular}{cccc}
$ K $    &   AV6       &  AV8      & RV8    \\
\hline
100      &   $-24.22$  &  $-25.15$ & $-23.35$       \\
200      &   $-24.90$  &  $-25.50$ & $-24.15$       \\
300      &   $-25.13$  &  $-25.60$ & $-24.35$       \\
400      &   $-25.25$  &  $-25.62$ & $-24.49$       
\end{tabular}
\end{table}
\begin{table}
\caption{Energies (in MeV) and radii (in fm) of the triton by different 
methods and with different interactions. The probability $P_L$ is 
given in \%.   }
\begin{tabular}{cccccc}
                  &   SVM     &  GFMC&   FY    & VMC & CHH    \\
\hline
AV6                           &           &            &       &       		&         \\
\hline
$\langle T   \rangle         $&$ 44.8$   &             &       &        	&         \\ 
$\langle V_6 \rangle         $&$-51.9$   &$-52.0(3.00)$&       &$-43.7(1.0)$      &         \\
$\langle r^2 \rangle^{1\over 2}$&$1.76$  &$1.75(0.10)$ &       &$1.95(0.03)$    	&         \\
$        P_S                 $&$91.2$    &             &       &        	&         \\
$        P_P                 $&$<0.1$     &             &       &      	  	&         \\
$        P_D                 $&$8.7$     &             &       &	        &         \\
$        E                   $&$-7.15$   &$-7.22(0.12)$\tablenotemark[1]&$-7.15$\tablenotemark[1]&$-6.33(0.05)$\tablenotemark[1]&   \\
\hline
AV8                           &           &             &       &       &          \\
\hline
$\langle T   \rangle         $&$ 46.3$   &             &       &        &         \\
$\langle V_6 \rangle         $&$-52.9$   &		&       &       &          \\
$\langle V_{LS} \rangle      $&$-1.2$    &             &       &        &         \\
$\langle r^2 \rangle^{1\over 2}$&$1.75$  &		&       &       &          \\
$        P_S                 $&$91.1$    &             &       &        &         \\
$        P_P                 $&$<0.1$   &             &       &        &         \\
$        P_D                 $&$8.9$     &             &       &        &         \\
$        E                   $&$-7.79$   &		&$-7.79\tablenotemark[4]$&  &$-7.79$\tablenotemark[3] \\
\hline
RV8                           &           &             &       &       &          \\
\hline
$\langle T   \rangle       $& $52.2$ &$54.0(0.20)$ &$52.2$   &         &         \\
$\langle V   \rangle       $& $-59.8$ &$-62.0(0.20)$&$-59.8$ &         &          \\
$\langle r^2 \rangle^{1\over 2}$&$1.75$     &$1.68(0.07)$ &$1.76$&        &          \\
$        P_S                 $& $ 90.3$     &             &       &        &         \\
$        P_P                 $& $<0.1$     &             &       &        &         \\
$        P_D                 $& $ 9.7$     &             &       &         &         \\
$        E                   $&$-7.59$    &$-7.54(0.10)$\tablenotemark[2]&$-7.59$\tablenotemark[2]&
$-7.44(0.03)$\tablenotemark[5]
&$-7.60$\tablenotemark[3]      
\end{tabular}
\tablenotetext[1]{Ref.\ \cite{GFMC1}.}
\tablenotetext[2]{Ref.\ \cite{GFMC2}.}
\tablenotetext[3]{Ref.\ \cite{CHH1}.}
\tablenotetext[4]{Ref.\ \cite{FY2}.}
\tablenotetext[5]{Ref.\ \cite{VMC1}.}
\end{table}

\begin{table}
\caption{Energies (in MeV) and radii (in fm) of the alpha particle by 
different methods and with different interactions. The probability $P_L$ is 
given in \%.  }
\begin{tabular}{cccccc}
                 &   SVM     &  GFMC             & FY  & VMC              & CHH       \\
\hline
AV6                          &            &              &       &     &          \\ 
\hline
$\langle T   \rangle         $& $ 100.1$   &              &        &                  &              \\
$\langle V_6 \rangle         $& $-125.4$  &$-122.0(3.0)$ &        &$-122.0(1.0)$     &         \\
$\langle r^2 \rangle^{1\over 2}$& $1.49 $  &$1.50(0.04)$  &        &$1.50(0.01)$     &         \\
$        P_S                 $& $84.3$  &              &        &     &         \\
$        P_P                 $& $0.5 $  &              &        &     &         \\
$        P_D                 $& $15.1$  &              &        &     &         \\
$        E                   $& $-25.25$  &$-24.79(0.20)$\tablenotemark[1]&        &$-22.75(0.01)$\tablenotemark[1]     &         \\     
\hline
AV8                             &           &              &        &       &         \\ 
\hline
$\langle T   \rangle         $  &$98.8$    & $$             &      &      &         \\
$\langle V \rangle           $  &$-124.4$  & $-124.20(1.0)$&        &      &          \\
$\langle V_6 \rangle         $  &$-121.5$  &                &        &      &          \\
$\langle V_{LS}\rangle       $  &$-2.9$    &                &        &     &         \\
$\langle r^2 \rangle^{1\over 2}$&$1.50$          & $1.51(0.01)$  &        &       &         \\
$        P_S                 $  & $85.5$         &              &        &       &         \\
$        P_P                 $  & $0.3$          &              &        &       &         \\
$        P_D                 $  & $14.2$          &              &        &       &         \\
$        E                    $  &$-25.62$   &$-25.75(0.02)$&$-25.31$\tablenotemark[3]&  &$-25.60$\tablenotemark[5] \\
\hline
RV8                               &           &              &        &      &         \\ 
\hline
$\langle T   \rangle             $&$111.7$   &$109.2(0.20)$ &        &     &         \\
$\langle V_6 \rangle             $&$-139.1$  &$-137.5(0.20)$&        &     &         \\
$\langle V_{LS}\rangle  $        &$2.1$      &$2.45(0.23)$  &        &     &         \\
$\langle V_{coul}\rangle  $      & $0.75$          &$0.71(0.02)$         &        &     &         \\
$\langle r^2 \rangle^{1\over 2} $& $1.51$     &$1.53(0.02)$  &        &     &         \\
$        P_S           $         & $84.1$    &              &        &     &         \\
$        P_P           $         & $0.4$     &              &        &     &         \\
$        P_D           $         & $15.5$    &$15.5(0.20)$   &        &     &         \\
$        E             $         &$-24.49$     &$-24.55(0.13)$\tablenotemark[2]&$-23.79$\tablenotemark[6]&$-23.26$\tablenotemark[4]&$-23.9$\tablenotemark[5]         
\end{tabular}
\tablenotetext[1]{Ref.\ \cite{GFMC1}.}
\tablenotetext[2]{Ref.\ \cite{GFMC2}.}
\tablenotetext[3]{Ref.\ \cite{FY2}.}
\tablenotetext[4]{Ref.\ \cite{VMC1}.}
\tablenotetext[5]{Ref.\ \cite{CHH1}.}
\tablenotetext[6]{Ref.\ \cite{FY1}.}
\end{table}

\end{document}